\documentclass[twocolumn,pra,showpacs,superscriptaddress]{revtex4}
\usepackage{amssymb}
\usepackage{amsmath}
\usepackage{graphicx}
\usepackage{subfigure}
\usepackage{natbib}
\usepackage{epsfig}
\usepackage{amsfonts}
\usepackage{mathrsfs}
\usepackage{ulem}
\usepackage{color}
\usepackage[toc,page,title,titletoc,header]{appendix}
\usepackage{CJK}
\usepackage{graphicx}

\begin{document}

\title{Interplay between Fano resonance and $\mathcal{PT} $ symmetry in non-Hermitian discrete systems}

\author{Baogang Zhu}
\affiliation{Beijing National
Laboratory for Condensed Matter Physics, Institute of Physics,
Chinese Academy of Sciences, Beijing 100190, China}
\author{Rong L\"u}
\affiliation{Department of Physics, Tsinghua University, Beijing 100084, China}
\affiliation{Collaborative Innovation Center of Quantum Matter, Beijing, China}
\author{Shu Chen}
\email{schen@aphy.iphy.ac.cn} \affiliation{Beijing National
Laboratory for Condensed Matter Physics, Institute of Physics,
Chinese Academy of Sciences, Beijing 100190, China}
\affiliation{Collaborative Innovation Center of Quantum Matter, Beijing, China}

\begin{abstract}
We study the effect of $\mathcal{PT} $-symmetric complex potentials on the transport properties of non-Hermitian systems, which consist of an infinite linear chain and two side-coupled defect points with $\mathcal{PT} $-symmetric complex on-site potentials.
By analytically solving the scattering problem of two typical models, which display standard Fano resonances in the absence of non-Hermitian terms,
we find that the $\mathcal{PT} $-symmetric imaginary potentials can lead to some pronounced effects on transport properties of our systems, including changes from the perfect reflection to perfect transmission, and rich behaviors for the absence or existence of the prefect reflection at one and two resonant frequencies. Our study can help us to understand the interplay between the Fano resonance and $\mathcal{PT}$ symmetry in non-Hermitian discrete systems, which may be realizable in optical waveguide experiments.
\end{abstract}


\maketitle
\date{today}

\section{Introduction}

Two decades ago Bender and Boettcher have found that a broad family of non-Hermitian Hamiltonians can exhibit
entirely real spectra as long as these Hamiltonians have parity-time ($\mathcal{PT} $) symmetry \cite{Bender}.
One distinguishing feature of $\mathcal{PT} $-symmetric Hamiltonians is the existence of spontaneous symmetry
breaking, corresponding to a transition from real to complex spectra \cite{Bender07}.
Since then, numerious $\mathcal{PT} $-symmetric systems have been explored in several fields, from the complex
extension of quantum mechanics \cite{quantum1,quantum2}, to the quantum field theories and mathematical physics \cite{Bender04}, open quantum systems \cite{Rotter09}, the Anderson models for disorder systems \cite{Goldsheid98,Heinrichs01,Molinari09}, the optical systems with complex refractive indices \cite{Klaiman08,Sukhorukov10,Ramezani12,Longhi09,Musslimani08,Luo13}, and the topological insulators \cite{Hu11,Zhubg}.

In recent years, the non-Hermitian lattice models with $\mathcal{PT} $ symmetry have been extensively studied, which is stimulated by
their experimental realizations in optical waveguides \cite{Guo09,Ruter10}, optical lattices \cite{Regensburger13}, and in a pair of coupled \emph{LRC} circuits \cite{Schindler11}. It is well known that the lattice models can exhibit rich physical phenomena due to the availability of exact solutions and the tractability of numerical and analytical calculations. Quiet recently, the non-Hermitian lattice models with $\mathcal{PT} $ symmetry have been studied in different systems, such as Gegenbauer-polynomial quantum chain \cite{Znojil10}, one-dimensional $\mathcal{PT} $-symmetric chain with disorder \cite{Bendix09}, the chain model with two conjugated imaginary potentials at two end sites \cite{Jin09}, the tight-binding model with position-dependent hopping amplitude \cite{Joglekar11}, and the time-periodic $\mathcal{PT} $-symmetric lattice model \cite{Valle13}. It is noted that many works have focused on the $\mathcal{PT} $ phase diagram and signatures of $\mathcal{PT} $-symmetry breaking in the lattice model, while much less attention has been paid to the transport properties \cite{Longhi13,Yogesh} and the associated quantum phase interference effects in the non-Hermitian lattice model with $\mathcal{PT} $ symmetry.

The transport properties of simple one- and quasi-one-dimensional model systems have been investigated with keen interest over the past two decades, which is due to the rapid advance of nanofabrication and detection techniques. One of the interesting phenomena in novel transport properties of such systems is the Fano resonance \cite{Fano}, which is characterized as a sharp asymmetric profile in transmission line, arising from the quantum interference between the discrete energy level and the continuum spectrum (for a recent review, see Ref. \cite{Miroshnichenko_RMP}). Interest in studying the Fano resonance is driven by its high sensitivity to the details of the scattering process, which shows the information of the scattering center.

The simplest model which reveals the underlying physics of coupling between the discrete states and the continuum spectrum is the Fano-Anderson model \cite{FanoAnderson,Anderson}. This model consists of a subsystem of an infinite linear main chain with the nearest-neighbor hopping which fertilizes the continuum spectrum, and a subsystem of several side-coupled discrete energy defects. These two subsystems are coupled at some joint sites by the hopping. Many attempts have been made to affect the scattering process and tune the transmission lineshape, including trying different geometric structure of defects \cite{CPB}, changing the numbers of defects, adding the impurities \cite{impurity}, or introducing the nonlinearity \cite{nonlinear1,nonlinear2,Andery1} to the subsystems. As another kind of interaction factor, the non-Hermitian terms may provide an interesting method to control and engineer the transmission lineshapes \cite{SongZhi,Andery1,Song2014}.

The aim of this paper is to study the transmission properties and the associated interference effect in two kinds of generalized non-Hermitian Fano-Anderson models. The models consist of an infinite linear main chain and two side-coupled defects with the $\mathcal{PT} $-symmetric complex on-site chemical potentials.
It is found that these systems exhibit typical Fano resonance in the Hermitian case, and the perfect reflection appears when the incoming frequency equals to the resonance energy, which is accompanied by an abrupt $\pi $-jump of the scattering phase \cite{Andery1}. While in the case of $\mathcal{PT} $-symmetric complex chemical potential, the transmission lineshape is shifted and tuned, and perfect transmission occurs when the the incoming frequency equals to the real part of the complex potential of the defects. Particularly in model (b) (as shown in Fig.1), we find that the resonance energy for perfect reflection is exact the real part of the complex potential, which indicates the perfect reflection in the absence of non-Hermitian term is changed to the perfect transmission at this frequency when the $\mathcal{PT} $-symmetric non-Hermitian potentials is acting. We show that the summation of transmission and reflection coefficients equals to $1$ exactly under the $\mathcal{PT} $-symmetric condition, i. e., the scattering process is similar to a Hermitian system due to the balanced gain and loss. The transport properties under the non-$\mathcal{PT} $-symmetric conditions are also calculated numerically, and the results show that the asymmetric Fano-type transmission profile is destroyed and the transmittance (reflectivity) can be larger than 1 as a result of imbalance of gain and loss, which indicates the instability of the process.

The paper is organized as follows. In Sec. \uppercase\expandafter{\romannumeral2}, we present the generalized non-Hermitian Fano-Anderson models and show the Fano resonance in a Hermitian version. In Sec. \uppercase\expandafter{\romannumeral3}, we study the scattering process of these two models, obtain the analytical form of transmission coefficient and discuss the interplay between $\mathcal{PT} $ symmetry and Fano resonance. A summary is given in Sec. \uppercase\expandafter{\romannumeral4}.

\section{Models}
We consider the generalized non-Hermitian Fano-Anderson model which consists of a linear main chain and two side-coupled defects with $\mathcal{PT} $-symmetric complex on-site chemical potentials. As schematically displayed in Fig. 1(a) and Fig. 1(b), two typical models are studied in the present work, which differ from each other by the kind of couplings to the main chain.
The model shown in Fig. 1(a) can be described by the Hamiltonian
\begin{figure}[!htb]
\includegraphics[width=8cm,]{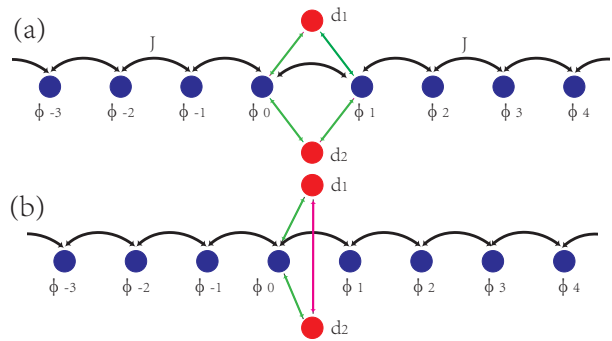}
\caption{(Color online) Schematic diagram of two kinds of generalized non-Hermitian Fano-Anderson models. The array of blue circles is the linear main chain, which contributes a continuum spectrum, and the isolated red circles are two defects with $\mathcal{PT} $-symmetric complex on-site chemical potentials. The couplings between different states are indicated by the black arrows.}
\end{figure}
\begin{eqnarray}
H_a &=&\underset{n}{\sum }J\hat{\phi} _{n-1}^{\dagger } \hat{\phi}_{n}+J_{\Vert }(\hat{d}_{1}^{\dagger }\hat{\phi}
_{0}+\hat{\phi}_{0}\hat{d}_{2}^{\dagger}+ \hat{d}_{1}^{\dagger}\hat{\phi}_{1} +\hat{d}_{2}^{\dagger } \hat{\phi}_{1}) \nonumber \\
&&+h.c.+(E_{d}+i\gamma ) \hat{d}_{1}^{\dagger} \hat{d}_{1}  +(E_{d}-i\gamma)
\hat{d}_{2}^{\dagger} \hat{d}_{2}  , \label{Hamiltonian1}
\end{eqnarray}
and the model in Fig. 1(b) can be described by the Hamiltonian
\begin{eqnarray}
H_b &=& \underset{n}{\sum }J\hat{\phi} _{n-1}^{\dagger } \hat{\phi}_{n} + J_{1 }\hat{d}_{1}^{\dagger } \hat{\phi}
_{0}+J_{2 }\hat{d}_{2}^{\dagger} \hat{\phi} _{0}+J_{\bot} \hat{d}_{2}^{\dagger}\hat{d}_{1}+ h.c. \nonumber \\
&&+(E_{d}+i\gamma ) \hat{d}_{1}^{\dagger} \hat{d}_{1}  +(E_{d}-i\gamma) \hat{d}_{2}^{\dagger} \hat{d}_{2} \label{Hamiltonian2}
\end{eqnarray}
where $\hat{\phi}_{n}$ ($\hat{\phi}_{n}^{\dagger}$) denotes the annihilation (creation) operator for annihilating (creating) a mode state $|\phi_{n} \rangle $ at site $n$ and $\hat{d}_{1,2}$ ($\hat{d}_{1,2}^\dagger$) the annihilation (creation) operator for annihilating (creating) a local mode  at the side site $d_{1,2}$. These two models describe the interaction of two subsystems. One is an infinite isotropic tight-binding chain with amplitude $\phi_{n} $ at site $n$, and $J $ is the nearest-neighbor hopping. This subsystem provides the fluent channel of the continuum spectrum for the propagation of plane waves with dispersion $\omega = 2J\cos k $. The other subsystem consists of two side-coupled defects with complex on-site chemical potentials, providing additional pathes for propagation. These two defects interact with the main chain at joint site $n=0$ and $1$ with the hopping amplitude $J_{\Vert}$ in the model (a) or at site $n=0$ only with the hopping amplitude $J_{1(2)} $ in the model (b).

In discrete systems, $\mathcal{P} $ and $\mathcal{T} $ are defined as the space-reflection (parity) operator and the time-reversal operator.
A Hamiltonian is said to be $\mathcal{PT} $ symmetric if it obeys the commutation relation $[\mathcal{PT},H]=0 $. In these two models the effect of $\mathcal{P} $ operator is $\mathcal{P} \hat{d}_{1}\mathcal{P}= \hat{d}_{2}$  ($\mathcal{P} \hat{d}_{2}\mathcal{P}= \hat{d}_{1}$) with the linear chain as the mirror axis, and the effect of $\mathcal{T} $ operator is $\mathcal{T}i \mathcal{T}= -i $. So it's easy to show that the Hamiltonian of model $(a)$ is invariant under the combined operation $\mathcal{PT} $, i.e., it's $\mathcal{PT} $-symmetric. As for the Hamiltonian of model $(b)$, to guarantee the $\mathcal{PT} $ symmetry, we require $J_{1}=J_{2}$, $E_{d1}=E_{d2}$ and $\gamma_{1} = -\gamma_{2}$.

To illustrate the typical Fano lineshape profile, we first consider the Hermitian case with $\gamma_{1(2)}=0$ in model $H_{b}$. For the case with only one side-coupled defect, i.e., $J_{2(\bot)}=E_{d2}=0$, which reduces to the standard single-level Fano-Anderson model \cite{FanoAnderson}, the transmission coefficient $T$ can be represented in the form
\begin{equation}
T=\frac{(\omega -E_{d1})^{2}}{(\omega -E_{d1})^{2}+(\frac{J_{1}^{2}}{%
2J\sin k})^{2}}=\frac{(\alpha _{k}+q)^{2}}{\alpha _{k}^{2}+1}, \label{OneSideCoupled}
\end{equation}
where $\alpha_{k}=2J \sin k(\omega -E_{d1})/ J_{1}^{2}$, in which $\omega=2J\cos k $ is the plane wave frequency, $q$ is the asymmetry parameter or Fano factor (in this case $q=0$), and $E_{d1}$ is the resonance energy. The right part of Eq. (\ref{OneSideCoupled}) is definitely the Fano formula. Perfect reflection occurs when the incoming frequency resonates with the discrete defect energy, i.e., $T=0$ at $\omega =E_{d1}$, as shown in Fig. 2(a). As is known, there is a $\pi$-jump of the scattering phase at the resonance \cite{Andery1}, which reveals the origin of the destructive interference. Fig. 2(b) shows the phase jump at the resonance energy $\omega=E_{d1}$. Obviously, the incoming wave frequency $\omega =2J\cos k $ is in the interval [-2J,2J] for arbitrary incoming $k$, so the requirement for the resonant scattering is that the discrete energy must be in the region of the continuum spectrum, i.e., $\left\vert E_{d1} \right\vert \leq 2J $. And the case of two defects in Hermitian case is a generalization of the case of one defect, with $T$ in the form
\begin{equation}
T=\frac{(\omega -E_{d}-J_{\bot})^{2}}{(\omega -E_{d}-J_{\bot})^{2}+(\frac{J_{\Vert}^{2}}{%
J\sin k})^{2}} \label{twoSideCoupled}
\end{equation}
in which $J_{1}=J_{2}=J_{\Vert}$ and $E_{d1}=E_{d2}=E_{d} $. It shows similar phenomena with the one defect case with resonance energy $\omega=E_{d}+J_{\bot}$, as shown in Fig. 2(c) and (d). The red solid line stands for the case $J_{\bot}=0$, and the blue dash line $J_{\bot}=0.2$. In Fig. 2(e) and (f) we show the transmission of the model $(a)$, which has similar feature with the model $(b)$ with resonance energy at $\omega=E_{d}-2J_{\Vert}^2 /J$.
We note that the case with more complicated side-coupled defect structures was also found to demonstrate a similar phenomena with frequencies of perfect reflections occurring at the eigenmode frequencies of the isolated defect structures \cite{nonlinear1}.
\begin{figure}[!htb]
\includegraphics[width=9cm,]{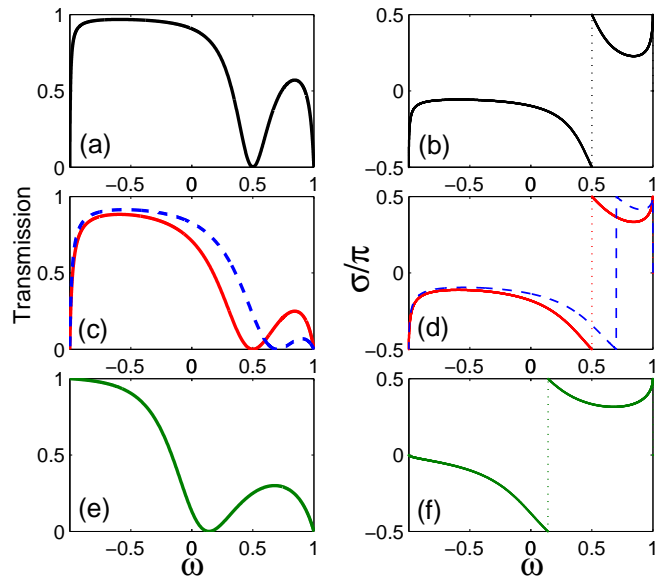}
\caption{(Color online) Transmission for structures with real chemical potentials. (a) and (b) are transmission coefficient T and scattering phase $\sigma$ for $H_{b}$ with only one defect at site 0, with parameters $J=0.5, J_{1}=0.4, E_{d1}=0.5$. (c) and (d) are for $H_{b} $ with two defects, with $J=0.5, J_{1}=J_{2}=0.4, E_{d1}= E_{d2}=0.5$. The red solid line stands for the case $J_{\bot}=0$, and the blue dash line for $J_{\bot}=0.2$. (e) and (f) are for $H_{a} $ with $J=0.5$, $J_{\Vert}=0.3$ and $E_{d}=0.5$.}
\end{figure}

\section{Non-Hermitian $\mathcal{PT} $-Symmetric System}

In this section, we study the transport properties of the non-Hermitian $\mathcal{PT}$-symmetric system.
We derive the analytical formula of the transmission coefficient with the help of the transfer-matrix method.
If we expand the wave function of the system
as $|\psi \rangle = \sum_{n} \phi_{n}(\tau) |\phi_n \rangle + d_{1}(\tau)|d_{1} \rangle  + d_{2}(\tau)|d_{2} \rangle$ with $|\phi_n \rangle=\hat{\phi}_n^{\dagger} |0\rangle$
and $|d_{1,2} \rangle =\hat{d}_{1,2}^{\dagger} |0\rangle$, from the equation $i\partial_{\tau} |\psi \rangle = H_a |\psi \rangle$, the following coupled-mode equations can be derived for the expansion coefficients $\phi_{n}(\tau)$ and $d_{1,2}(\tau)$:
\begin{eqnarray*}
i\dot{\phi}_{n} &=&J\phi _{n-1}+J\phi _{n+1}+J_{\Vert }(d_{1}+d_{2})\delta
_{n0}\\
      & & +J_{\Vert }(d_{1}+d_{2})\delta _{n1}, \\
i\dot{d}_{1} &=&(E_{d}+i\gamma )d_{1}+J_{\Vert }\phi _{0}+J_{\Vert }\phi _{1},
\\
i\dot{d}_{2} &=&(E_{d}-i\gamma )d_{2}+J_{\Vert }\phi _{0}+J_{\Vert }\phi _{1},
\end{eqnarray*}
where the overdot stands for the derivative of time $\tau$. The stationary solution can be expressed as the following form
\begin{equation*}
\phi _{n}(\tau)=A_{n}e^{-i\omega \tau },d_{1(2)}(\tau )=B_{1(2)}e^{-i\omega
\tau }
\end{equation*}
and then we obtain the algebraic relationship of the amplitudes on each site, the following equations indicate that they are nested with each other
\begin{eqnarray*}
\omega A_{n} &=&JA_{n-1}+JA_{n+1}+J_{\Vert }(B_{1}+B_{2})\delta
_{n0}\\
        & & +J_{\Vert }(B_{1}+B_{2})\delta _{n1}, \\
\omega B_{1} &=&(E_{d}+i\gamma )B_{1}+J_{\Vert }A_{0}+J_{\Vert }A_{1}, \\
\omega B_{2} &=&(E_{d}-i\gamma )B_{2}+J_{\Vert }A_{0}+J_{\Vert }A_{1}.
\end{eqnarray*}
We can easily derive the form of $B_{1} $ and $B_{2} $ as a function of $A_{0} $ and $A_{1} $, and substitute it into the first line formula to eliminate two degrees
\begin{eqnarray*}
B_{1}=J_{\Vert }\frac{A_{0}+A_{1}}{\omega -(E_{d}+i\gamma )},B_{2}=J_{\Vert }
\frac{A_{0}+A_{1}}{\omega -(E_{d}-i\gamma )}
\end{eqnarray*}
and obtain the effective defect equation
\begin{eqnarray}
\omega A_{n} &=& JA_{n-1}+JA_{n+1}+F_{d}(A_{0}+A_{1})\delta
_{n0} \nonumber \\
&& +F_{d}(A_{0}+A_{1})\delta _{n1} ,
\end{eqnarray}
in which
\begin{eqnarray*}
F_{d}=2J_{\Vert }^{2}\frac{\omega -E_{d}}{(\omega -E_{d})^{2}+\gamma ^{2}}
\end{eqnarray*}
is an entirely real function acting as an effective localized potential. This equation reveals that this problem can be simplified to the propagating problem along the main chain with scattering center at site 0 and 1, and the defect subsystem acts as the effective localized potential. When the module of this effective localized potential is zero, the perfect transmission occurs, and correspondingly when it is infinite, the perfect reflection appears. For the scattering problem, the wave function can be expressed in the form
\begin{equation*}
A_{n}=\left\{
\begin{array}{c}
Ie^{ikn}+re^{-ikn},n < 0, \\
te^{ikn},n > 1,%
\end{array}%
\right.
\end{equation*}
where $I$, $r$, and $t$ stand for the amplitude of the incoming, reflected and transmitted wave, respectively, and $k $ is assumed to be positive, indicating the left incidence. The momentum $k$ is related to the $\omega$ by the dispersion equation $\omega=2J \cos k$. By substituting the expression $A_n$ into the effective defect equation  and applying the transfer-matrix method, we obtain the analytical form of the transmission amplitude
\begin{equation}
t = I \frac{i\sin k[F_d+J]}{F_d (1+e^{ik})+iJ\sin k} \label{t},
\end{equation}
and the transmission coefficient
\begin{equation}
T = \left\vert t\right\vert ^{2}/I^{2} .
\end{equation}
Obviously, the condition for the perfect reflection $T=0$ is $F_{d}+J=0 $, which leads to two solutions of $\omega$
\begin{equation}
\omega =E_{d}-\frac{J_{\Vert }^{2}\pm \sqrt{J_{\Vert
}^{4}-J^{2}\gamma ^{2}}}{J} , \label{omegat0}
\end{equation}
as long as $\gamma$ is smaller than a critical value $\gamma_{c}= J_{\Vert }^{2}/J $.
\begin{figure}[!htb]
\includegraphics[width=9cm,]{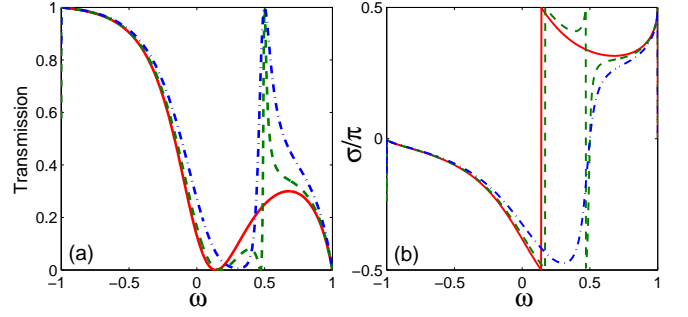}
\caption{(Color online) (a) Transmission coefficient for non-Hermitian $\mathcal{PT} $-symmetric model $H_{a}$ with different $\gamma$: the red solid line stands for $\gamma=0$, the green dash one for $\gamma=0.1$, and the blue dot-dash one for $\gamma=0.2$. (b) indicate the scattering phase $\sigma/\pi$ for the corresponding condition. Here $J=0.5$, $J_{\Vert}=0.3$ and $E_{d}=0.5$.}
\end{figure}

In Fig. 3, we show the scattering results of the model $(a)$ for different $\gamma$. When $\gamma=0 $, the condition has be discussed above with only one resonance energy at $\omega=E_{d} - 2\gamma_{c}$. The scattering phase $\sigma= \arg (t)$ shows a $\pi$-jump correspondingly, shown as the red solid line in Fig. 3(b). While for $\gamma=0.1$, which is smaller than $\gamma_{c}=0.18$, there exists two resonance energies determined by Eq. (\ref{omegat0}), at which the transmission is completely suppressed and the scattering phase $\sigma$ experiences an abrupt $\pi$-jump or $-\pi$-jump. What's interesting is that at $\omega=E_{d}$, $t = 1$ due to $F_{d}\equiv 0$. It means that a perfect transmission occurs at the energy $E_{d}$, which is caused by the balanced non-Hermitian terms $\pm i \gamma$, shown as the green dash line in Fig. 3. When $\gamma=0.2$, which is larger than $\gamma_{c}$, there exists no solution for the perfect reflection, and the scattering phase $\sigma$ experiences no abrupt jump. But the perfect transmission is still found for arbitrary $\gamma$, shown as the blue dot-dash line in Fig. 3. The formula of reflection coefficient $R=\left\vert r \right\vert ^{2}/I^{2}$ can be obtained by the similar method, and it's found that the summation of transmission and reflection coefficients equals to $1$ for arbitrary $\gamma$ value, i.e., $R+T\equiv 1$. That means this non-Hermitian system acts as a Hermitian one in the scattering process because of $\mathcal{PT} $ symmetry \cite{SongZhi}.

\begin{figure}[!htb]
\includegraphics[width=9cm,]{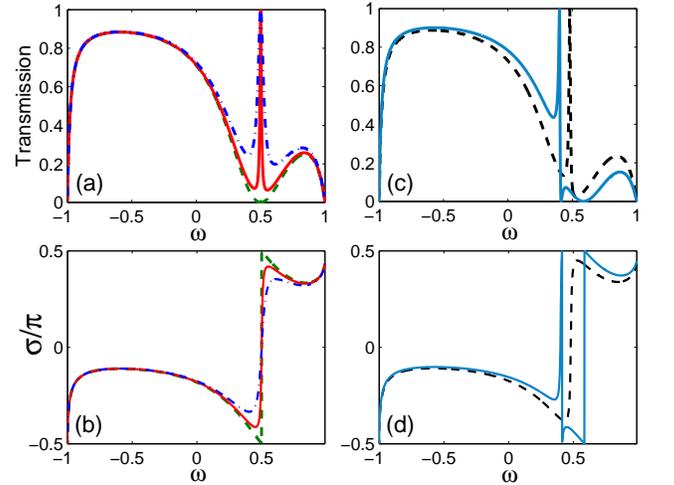}
\caption{(Color online) (a) Transmission coefficient for non-Hermitian $\mathcal{PT} $-symmetric model $H_{b}$ for different $\gamma$ with $J_{\bot}=0$: the green dash one stands for $\gamma=0$, the red solid one for $\gamma=0.05$, and the blue dot-dash one for $\gamma=0.1$. (b) indicates the scattering phase $\sigma/\pi$ for the corresponding system in (a). (c) Transmission coefficient for different $J_{\bot} $ with $\gamma=0.05$: the black dash line stands for $J_{\bot}=0.02$ and the blue dot-dash one for $J_{\bot}=0.1$. (d) indicates the scattering phase for the corresponding system in (c). Here $J=0.5$, $J_{\Vert}=0.4$ and $E_{d}=0.5$.}
\end{figure}

As for the model described by $H_{b}$, it can be related to the model $(a)$ by breaking the couplings between site $d_{1}$, $d_{2}$ and site 1, and introducing a coupling $J_{\bot}$ between $d_{1}$ and $d_{2}$. Nevertheless, it shows some special features due to its specific geometric structure. Following similar procedures as dealing with the model $(a)$,
we can obtain the effective defect equation
\begin{equation}
\omega A_{n}=JA_{n-1}+JA_{n+1}+\tilde{F}_{d}A_{0}\delta
_{n0}   \label{edeq2}
\end{equation}
with
\begin{eqnarray*}
\tilde{F}_{d}=\frac{[\omega -(E_{d_1}+i \gamma_1)]J_2^2+[(\omega -(E_{d_2}+i\gamma_2)]J_1^2+2J_1 J_2 J_{\bot}}{[\omega -(E_{d_1}+i \gamma_1)][(\omega -(E_{d_2}+i\gamma_2)]- J_{\bot}^{2}}.
\end{eqnarray*}
Then we can derive the transmission amplitude
\begin{equation}
t = I \frac{ i 2J \sin k}{  \tilde{F}_d + i 2J\sin k} \label{tb},
\end{equation}
in the scheme of transfer-matrix method.
Particularly, for the $\mathcal{PT}$-symmetrical model with $J_1=J_2=J_{\Vert}$, $E_{d_1} = E_{d_1}=E_{d}$ and $\gamma_1=- \gamma_2 = \gamma$, the effective defect potential $\tilde{F}_{d}$ becomes real with the form of
\begin{eqnarray*}
\tilde{F}_{d}=2J_{\Vert}^2 \frac{\omega -E_{d}+ J_{\bot}}{(\omega -E_{d})^2+ \gamma^2- J_{\bot}^{2}}
\end{eqnarray*}
and the transmission coefficient $T$ reduces to
\begin{eqnarray}
T =\frac{[(\omega -E_{d})^{2}+\gamma ^{2}-J_{\bot}^{2}]^{2}}{[(\omega
-E_{d})^{2}+\gamma ^{2}-J_{\bot}^{2}]^{2}+[\frac{J_{\Vert }^{2}}{J \sin
k}(\omega -E_{d}+ J_{\bot})]^{2}}.  \label{T-b}
\end{eqnarray}
When $\gamma=0 $, $T$ reduces back to Eq. (\ref{twoSideCoupled}) discussed above in Sec. \uppercase\expandafter{\romannumeral2} with a perfect reflection occurring at $\omega=E_{d}+J_{\bot}$ and an abrupt $\pi$-jump of the scattering phase $\sigma=\arg(t)$.

Now we discuss the effect of the $\mathcal{PT} $-symmetrical non-Hermitian terms.  We first consider the case with no interaction between $d_{1}$ and $d_{2}$, i.e., $J_{\bot}=0$. From Eq. (\ref{T-b}), we can see no perfect reflection occurring due to the existence of a nonzero $\gamma $. Instead, a perfect transmission occurs at $\omega=E_{d}$, as shown in Fig. 4(a) by the red solid ($\gamma=0.05$) and blue dot-dash ($\gamma=0.1$) line. This is the most pronounced feature different from the Fano profile of a Hermitian system. Correspondingly, the scattering phase $\sigma $ also indicates the deviation of $\pi $-jump as shown in Fig. 4(b). For the case with a nonzero $J_{\bot}$ term, from Eq. (\ref{T-b}), we obtain $T=1$ when $\omega=E_{d}-J_{\bot}$, i.e, the frequency for occurring a perfect transmission is shifted to $E_{d}-J_{\bot}$. Also, it's easy to show $T=0$ when
\begin{equation}
\omega=E_{d}\pm \sqrt{J_{\bot}^{2} - \gamma ^{2}},
\end{equation}
as long as $|J_{\bot}| >|\gamma|$. In Fig. 4(c) and Fig. 4(d) we plot the transmission coefficient and the associated scattering phase as a function of $\omega$ for different $J_{\bot}$. The case of $J_{\bot}=0.02$ shows no perfect reflection as it is smaller than $\gamma=0.05 $, and the corresponding scattering phase shows no $\pi$-jump, shown as the black dash line. But for the case of $J_{\bot}=0.1$ ($J_{\bot}> \gamma $), the line shape of transmission has two zero points and the scattering phase exhibits a $\pi$-jump meanwhile, shown as the blue solid line. We can also obtain the explicit form of the reflection coefficient as
\begin{eqnarray}
R =\frac{[\frac{J_{\Vert }^{2}}{J \sin
k}(J_{\bot}+\omega -E_{d})]^{2}}{[(\omega
-E_{d})^{2}+\gamma ^{2}-J_{\bot}^{2}]^{2}+[\frac{J_{\Vert }^{2}}{J \sin
k}(\omega -E_{d}+J_{\bot})]^{2}} . \label{R-b}
\end{eqnarray}
From Eq.(\ref{T-b}) and Eq.(\ref{R-b}), it's quite obvious that $R+T = 1$ for arbitrary value of $\gamma$, which indicates the $\mathcal{PT} $-symmetric non-Hermitian system having a similar behavior as a Hermitian one in the scattering process \cite{SongZhi}. We can also get the the explicit form of the reflection coefficient in model $H_{a} $ as
\begin{equation}
r=-I\frac{F_{d}(1+e^{ik}+ie^{ik}\sin k)}{F_{d}(1+e^{ik})+iJ\sin k}
\end{equation}
and then one can check $R+T = 1$ for arbitrary value of $\gamma$.
\begin{figure}[!htb]
\includegraphics[width=9cm,]{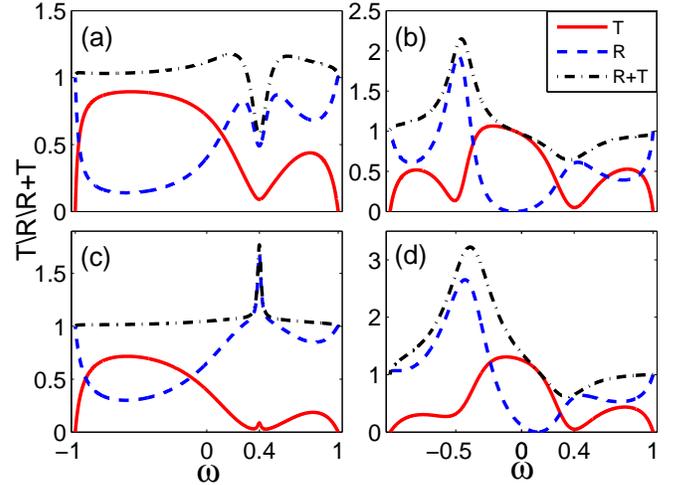}
\caption{(Color online) Transmission coefficient for the non-$\mathcal{PT} $-symmetric case of $H_b$. The red solid line represents $T$, the blue dash line represents $R$ and the black dot-dash line represents $R+T$. Parameters are $J=0.5$, (a) $J_{1}=J_{2}=0.4, E_{d1}=E_{d2}=0.4$, but $\gamma_{1}=0.05, \gamma_{2}=-0.15$. (b) $J_{1}=J_{2}=0.4, \gamma_{1}=-\gamma_{2}=0.05 $, but $E_{d1}=0.4, E_{d2}=-0.5$. (c) $E_{d1}=E_{d2}=0.4, \gamma_{1}=-\gamma_{2}=0.05 $, but $J_{1}=0.4, J_{2}=0.6$. (d) All parameters do not obey balanced rules, $J_{1}=0.4, J_{2}=0.6, E_{d1}=0.4, E_{d2}=-0.5, \gamma_{1}=0.05, \gamma_{2}=-0.15$.}
\end{figure}

To investigate the important role of $\mathcal{PT} $ symmetry, finally we consider several example cases by breaking the $\mathcal{PT} $ symmetry.
In order to compare with its $\mathcal{PT} $-symmetric correspondence, we consider the model $H_{b}$ by deviating the $\mathcal{PT} $-symmetrical conditions: $J_{1}=J_{2}$, $E_{d1}=E_{d2}$ and $\gamma_{1} = -\gamma_{2}$. In Fig. 5 we show several imbalanced cases for the model $H_{b}$. The transmission coefficient $T $ and reflection coefficient $R $ can be larger than $1$ depending on the strength of the gain and loss, which will lead to instability. The numerical results clearly show that $R+T$ is no longer a unit constant. Alternatively, one can also understand this from the effective defect equation described by Eq. (\ref{edeq2}), in which the effective defect potential $\tilde{F}_{d}$ is generally complex and becomes real only under the $\mathcal{PT}$-symmetric condition.
\begin{figure}[!htb]
\includegraphics[width=9 cm]{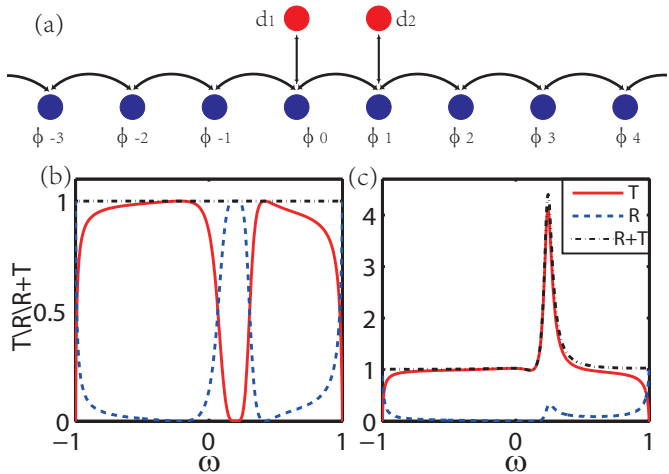}
\caption{(Color online) (a) Schematic diagram of another $\mathcal{PT} $ symmetric model. (b) Transmission and
reflection coefficient for the Hermitian case with $\gamma=0 $. (c) is for the non-Hermitian case with
$\gamma=0.1$. The red solid line represents T, the blue dash line represents R and the black
dot-dash line represents R+T. Parameters are $J=0.5, J_{\bot}=0.3, E_{d}=0.2$.}
\end{figure}

As we have already displayed that $R+T=1$ holds true for both the model (a) and (b) as long as the $\mathcal{PT} $ symmetry exists, it is interesting to ask whether the property of $R+T=1$ is a general feature of one-dimensional systems with $\mathcal{PT} $ symmetry? As a general proof is lack, we would like to indicate that this property is not general to all the non-Hermitian systems with $\mathcal{PT} $ symmetry by considering an example model described by
\begin{eqnarray}
H &=&\underset{n}{\sum }J\hat{\phi}_{n-1}^{\dagger }\hat{\phi}_{n}+J_{\bot }%
\hat{d}_{1}^{\dagger }\hat{\phi}_{0}+J_{\bot }\hat{d}_{2}^{\dagger }\hat{\phi%
}_{1}+h.c. \nonumber \\
&&+(E_{d}+i\gamma )\hat{d}_{1}^{\dagger }\hat{d}_{1}+(E_{d}-i\gamma )\hat{d}%
_{2}^{\dagger }\hat{d}_{2},
\end{eqnarray}
which is schematically displayed in Fig. 6(a). This model is $\mathcal{PT} $ symmetric with the vertical line between site 0 and site 1 as the mirror axis. In this model the effect of $\mathcal{P} $ operator is $\mathcal{P} \hat{\phi}_{n}\mathcal{P}= \hat{\phi}_{-n+1}$ and $\mathcal{P} \hat{d}_{1}\mathcal{P}= \hat{d}_{2}$  ($\mathcal{P} \hat{d}_{2}\mathcal{P}= \hat{d}_{1}$), and the effect of $\mathcal{T} $ operator is $\mathcal{T}i \mathcal{T}= -i $.
Following previous procedures, we can derive
the transmission and reflection amplitudes as
\begin{eqnarray}
t &=&Ie^{-ik}\frac{-2iJ^{2}\sin k}{(G_{d}-Je^{-ik})(G_{d}^{\ast
}-Je^{-ik})-J^{2}}, \\
r &=&I\frac{J^{2}-(G_{d}-Je^{ik})(G_{d}^{\ast }-Je^{-ik})}{%
(G_{d}-Je^{-ik})(G_{d}^{\ast }-Je^{-ik})-J^{2}}
\end{eqnarray}
with
\begin{eqnarray*}
G_{d}=\frac{J_{\bot }^{2}}{\omega -(E_{d}+i\gamma )},
\end{eqnarray*}
and the transmission coefficient $T$ and reflection coefficient $R$ are given by
\begin{eqnarray}
T=\left\vert t\right\vert ^{2}/I^{2}, ~~ R=\left\vert r\right\vert ^{2}/I^{2} .
\end{eqnarray}
While we have $R+T=1$ for the Hermitian case with $\gamma=0$ as shown in Fig. 6(b), but the transmission coefficient can be larger than $1$ around the resonance energy $E_d$ when $\gamma$ is nonzero as shown in Fig. 6(c), which indicates that $R+T\neq 1$ in this regime.
Our numerical result shows that this model does not obey $R+T=1$ although it is a $\mathcal{PT}$-symmetric system. Other $\mathcal{PT}$-symmetric systems with $R+T \neq 1$ can be found in references \cite{Jones,Znojil}.

\section{Summary}
In summary, we study two kinds of discrete non-Hermitian models, consisting of an infinite linear chain and two defect points with $\mathcal{PT} $-symmetric complex on-site chemical potentials, which can be regarded as the generalization of the Fano-Anderson model. By solving the wave propagation problems, we derive analytical results for the transmission coefficients. In comparison with standard Fano resonances in the Hermitian models, our results show that the $\mathcal{PT} $-symmetric non-Hermitian models can go through perfect transmission when the incoming frequency $\omega $ resonates with some specific energies. Depending on the choice of parameters, perfect reflection, accompanying by the emergence of an abrupt $\pi$-jump of the scattering phase, may also appear at one resonant frequency or two resonant frequencies. The condition for the appearance of perfect reflection is also discussed. Under the $\mathcal{PT}$-symmetric condition, the summation of transmission coefficient and reflection coefficient always equals to the unit constantly, which means the conservation of probability. We also show that the conservation of probability is unstable against $\mathcal{PT} $-symmetry-breaking perturbations.

\section*{Acknowledgment}
 This work is supported by NSF of China under Grants No. 11425419, No. 11374354 and No. 11174360, and by National Program for Basic Research of MOST. R. L. is supported by the NSFC under Grant No. 11274195 and the National Basic Research Program of China (973 Program) Grant No. 2011CB606405 and No. 2013CB922000.

\end{document}